\documentclass[pre,twocolumn,aps,floatfix,10pt]{revtex4}
\usepackage{amsmath}
\usepackage{amsfonts}
\usepackage{amssymb}
\usepackage{graphicx}
\usepackage{fontenc}
\usepackage{psfrag,layout}
\usepackage{upgreek}

\begin{document}
\title{Crystal-liquid interfacial free energy via thermodynamic integration}
\author{Ronald Benjamin and J{\"u}rgen Horbach}
\affiliation{Institut f{\"u}r Theoretische Physik II: Soft Matter, 
Heinrich Heine-Universit{\"a}t D{\"u}sseldorf,
Universit\"atsstra\ss e 1, 40225 D{\"u}sseldorf, Germany}

\begin{abstract}
A novel thermodynamic integration (TI) scheme is presented to compute the
crystal-liquid interfacial free energy ($\gamma_{\rm cl}$) from molecular
dynamics simulation. The scheme is applied to a Lennard-Jones system.
By using extremely short-ranged and impenetrable Gaussian flat walls to
confine the liquid and crystal phases, we overcome hysteresis problems
of previous TI schemes that stem from the translational movement of
the crystal-liquid interface.  Our technique is applied to compute
$\gamma_{\rm cl}$  for the (100), (110) and (111) orientation of the
crystalline phase at three temperatures under coexistence conditions. For
one case, namely the (100) interface at the temperature $T=1.0$ (in
reduced units), we demonstrate that finite-size scaling in the framework
of capillary wave theory can be used to estimate $\gamma_{\rm cl}$
in the thermodynamic limit. Thereby, we show that our TI scheme is not
associated with the suppression of capillary wave fluctuations.
\end{abstract}

\maketitle

\section{Introduction}
\label{sec_intro}
The interfacial free energy between a crystal phase in coexistence
with its liquid phase, $\gamma_{\rm cl}$, is an important
thermodynamic quantity which controls the rate of homogeneous
nucleation and the morphology of a crystal growing from its
melt~\cite{turnbull-cech50,turnbull50,turnbull52,woodruff73,tiller91}.
The magnitude and anisotropy in $\gamma_{\rm cl}$ determines whether
crystal growth occurs in a planar, cellular or dendritic manner. Even
a small dependence on the orientation of the crystal, e.g., may
significantly affect the final microstructure of the growing crystal
and the materials properties of the resulting crystal. Moreover, the
crystal-liquid interfacial free energy also affects the barrier for the
formation of a nucleus near a wall as well as the wetting behavior of
the crystal.

In spite of its central importance  for a proper understanding of
growth and nucleation phenomena, reliable experimental data for
$\gamma_{\rm cl}$ is hard to come by. Most experimental studies
obtain the crystal-liquid interfacial free energy indirectly by
applying classical nucleation theory to homogeneous nucleation rate
measurements~\cite{turnbull-cech50,turnbull50,woodruff73,broughton-gilmer86,kelton91}.
Such nucleation rate measurements represent an average over the various
orientations of the crystal and therefore provide no information regarding
the anisotropy of the crystal. Besides, the assumptions inherent in
classical nucleation theory affect the estimates of $\gamma_{\rm cl}$.
With state-of-the-art experiments, however, $\gamma_{\rm cl}$ can
be estimated in certain parameter ranges with an accuracy of 10-20\%
\cite{bokeloh11,bokeloh14,palberg14}.

In recent years, molecular simulations have emerged as
a widely used tool to understand the thermodynamics of
crystal-liquid interfaces from a microscopic perspective and obtain
reliable estimates of the the crystal-liquid interfacial free energy
\cite{huitema99,tepper-briels2002,laird-hamet92,davlaird96,davlaird98,hoyt2001,morris2002,morris-song2003,davidchack2006,amini-laird2008,zykova2009,zykova2010,rozas2011,
hartel2012,heinonen2013,martin2012,finnis2010,uberti2011,davidchack-laird2000,davidchack-laird03,laird-davidchack2005prl,laird-davidchack2005jpcb,davidchack2010,turci2014_1,turci2014_2}
Currently, several independent simulation
techniques are available to determine the
crystal-liquid interfacial free energy.  In the capillary fluctuation
method~\cite{hoyt2001,morris2002,morris-song2003,davidchack2006,amini-laird2008,zykova2009,zykova2010,rozas2011,hartel2012,heinonen2013},
fluctuations in the height of the crystal-liquid interface are used to
compute the interfacial stiffness.  Then by using an expansion of the
anisotropic interfacial stiffness into cubic harmonics, one can extract,
albeit indirectly, $\gamma_{\rm cl}$.

Recently, two related methods to compute $\gamma_{\rm cl}$ have been
proposed by Fern{\'{a}}ndez {\it {et al.}}~\cite{martin2012} and
Angioletti-Uberti {\it {et al.}}~\cite{finnis2010,uberti2011}, known
as tethered Monte Carlo and metadynamics, respectively. These methods
render accurate estimates of $\gamma_{\rm cl}$ with relatively moderate
system sizes (less than $10^4$ particles).  However, both approaches
rely on a local order parameter which has been specified only for the
(100) orientation of the fcc crystal structure.

In contrast, thermodynamic integration (TI)~\cite{frenkel-smit02,straatsma93}
can yield a direct estimate of $\gamma_{\rm cl}$ for various crystal
structures, without the necessity of introducing local order parameters
and with system sizes of a few thousand atoms. The determination
of $\gamma_{\rm cl}$ from thermodynamic integration is based on the
definition that the crystal-liquid interfacial free energy is the
reversible work required to create a unit area of the crystal-liquid
interface~\cite{adamson97}. Hence, one needs to find a reversible path
bringing together independent liquid and crystal phases to create two
crystal-liquid interfaces in equilibrium with the bulk phases.

Broughton and Gilmer~\cite{broughton-gilmer86} first  introduced a
TI scheme to compute $\gamma_{\rm cl}$ for a modified Lennard-Jones
(LJ) potential. Their method consisted of using specially designed
``cleaving potentials" to gradually split each phase into two blocks
divided by a cleaving plane, then bringing these blocks together,
and finally removing the ``cleaving potential".  However, designing
``cleaving" potentials for various orientations of the crystal, at
different coexistence conditions and for different potentials is clearly
an arduous task and the ``cleaving potentials" proposed by them cannot
be easily generalized.  Moreover, their calculations were not precise
enough to resolve the anisotropy in $\gamma_{\rm cl}$.

Davidchack and
Laird~\cite{davidchack-laird2000,davidchack-laird03,laird-davidchack2005prl,laird-davidchack2005jpcb,davidchack2010}
modified the approach of Broughton and Gilmer by using planar ``cleaving
walls" made of similar particles as the system and consisting
of one or more crystalline layers with the same structure as the
actual crystal phase. Subsequently, several authors have adopted the
``cleaving walls" scheme to compute $\gamma_{\rm cl}$ for various model
systems~\cite{tipeev13,wang-apte2013,apte2010}.  The data reported by
them were precise enough to resolve the anisotropy in $\gamma_{\rm cl}$.
However, the TI paths corresponding to both the Broughton-Gilmer and
Davidchack-Laird approaches are subject to hysteresis in the final
step, when the ``cleaving walls'' or the ``cleaving potential''
is removed. Due to thermal fluctuations, the two crystal-liquid
interfaces can change their position by simultaneous freezing and
melting~\cite{davidchack-laird03,davidchack2010}.  As a result, their
location will no longer coincide with the position of the cleaving plane,
leading to hysteresis between the forward and reverse TI paths.

Some authors have tried to adopt the ``cleaving walls'' TI scheme of
Davidchack and Laird~\cite{davidchack-laird2000,davidchack-laird03} into
a non-equilibrium work measurement approach~\cite{musong2006}. Such
an approach is independent of the reversibility of the
transformation~\cite{jarzynski1997,jarzynski06}. However, the
non-equilibrium work measurement still requires one to be able to
reach the initial state from the final state when the transformations
are carried out in the reverse direction.  Due to the movement of
the crystal-liquid interface, this is not possible.  The hysteresis
arising from the mobility of the crystal-liquid interface seems to be
a difficult problem to eliminate completely.  Though, Davidchack and
Laird~\cite{davidchack-laird03,davidchack2010} tried to deal with this
issue by performing several independent TI runs and choosing the run
with the least hysteresis.

In this work, we propose a novel thermodynamic integration scheme to
compute $\gamma_{\rm cl}$ from molecular dynamics simulations. Our scheme
circumvents the key problem due to the mobility of the crystal-liquid
interface by the use of a very short ranged flat wall (modelled by
a Gaussian potential) to split the crystal and liquid phases. Such
a short-ranged wall demands that the integration of the equations
of motion is carried out with a very small time step.  However,
this disadvantage can be offset by the use of a multiple time step
scheme~\cite{frenkel-smit02}. The contribution of this short-ranged wall
to the final value of the interfacial free energy itself is negligible
(much less than the statistical errors and less than $0.1\%$ of the final
value of $\gamma_{\rm cl}$).  In addition to this short-ranged flat
wall, we also insert a structured solid wall consisting of frozen-in
crystalline layers, to gradually bring together in a smooth manner the
individual crystal and liquid phases split by the flat wall.  We employ
our thermodynamic integration scheme to compute $\gamma_{\rm cl}$ for
a modified Lennard-Jones potential, though our scheme can also be used
for more complex potentials. 

In order to determine $\gamma_{\rm cl}$ in the thermodynamic limit, we use
a careful analysis of finite-size effects in the framework of capillary
wave theory \cite{binder82,schmitz2014}. Although the finite-size effects
turn out to be relatively small for systems containing more than about
20000 particles, our finite-size analysis is nicely consistent with
the prediction of capillary-wave theory and thus this demonstrates
that our TI scheme does not lead to a suppression of capillary-wave
fluctuations. This indicates that the Gaussian flat walls introduced in
our scheme indeed lead to negligibly small perturbation of the system.

In the next section (Sec.~\ref{sec_model}), we introduce the Lennard-Jones
model for which $\gamma_{\rm cl}$ is computed. After describing our TI
scheme (Sec.~\ref{sec_ti}), the details of the simulation are given in
Sec.~\ref{sec_sim}.  The results are presented in Sec.~\ref{sec_res},
and finally, we end with a conclusion in Sec.~\ref{sec_conc}.

\section{Model}
\label{sec_model}
The Lennard-Jones (LJ) potential describes the interaction between a
particle $i$ and a particle $j$ separated by a distance $r_{ij}$ and is
given by
\begin{equation}
\phi(r_{ij}) =
4\epsilon 
\left[\left(\frac{\sigma}{r_{ij}}\right)^{12}
- \left(\frac{\sigma}{r_{ij}}\right)^{6} \right],
\label{eq:lj}
\end{equation}
where the parameters $\epsilon$ and $\sigma$ set the scales for energy and
length, respectively. Broughton and Gilmer~\cite{broughton-gilmer86}
proposed a modified LJ potential with a cut-off at $r^{\rm
cut}=2.5\,\sigma$ that provides continuity of potential and force at
$r=r^{\rm cut}$. It is defined by
\begin{equation}
u(r_{ij}) = \phi(r_{ij}) + C_{1} 
\label{eq:ur1}
\end{equation}
for $0<r_{ij}\leq2.3\,\sigma$,
\begin{equation}
 u(r_{ij}) = C_{2} \left(\frac{\sigma}{r_{ij}}\right)^{12} 
+ C_{3} \left(\frac{\sigma}{r_{ij}}\right)^{6}
+ C_{4}\left(\frac{r_{ij}}{\sigma}\right)^{2} + C_{5} 
\label{eq:ur2}
\end{equation}
for $2.3\,\sigma<r_{ij}<r^{\rm cut}=2.5\,\sigma$ and $u(r_{ij})=0$
for $r_{ij}\geq r^{\rm cut}$. The constants in Eqs.~(\ref{eq:ur1})
and (\ref{eq:ur2}) are given by $C_{1}=0.016132\,\epsilon$,
$C_{2}=3136.6\,\epsilon$, $C_{3}=-68.069\,\epsilon$,
$C_{4}=-0.083312\,\epsilon$, and $C_{5}=0.74689\,\epsilon$.

\section{Thermodynamic Integration Scheme}
\label{sec_ti}
\begin{figure}
\includegraphics[width=3.0in,scale=0.5]{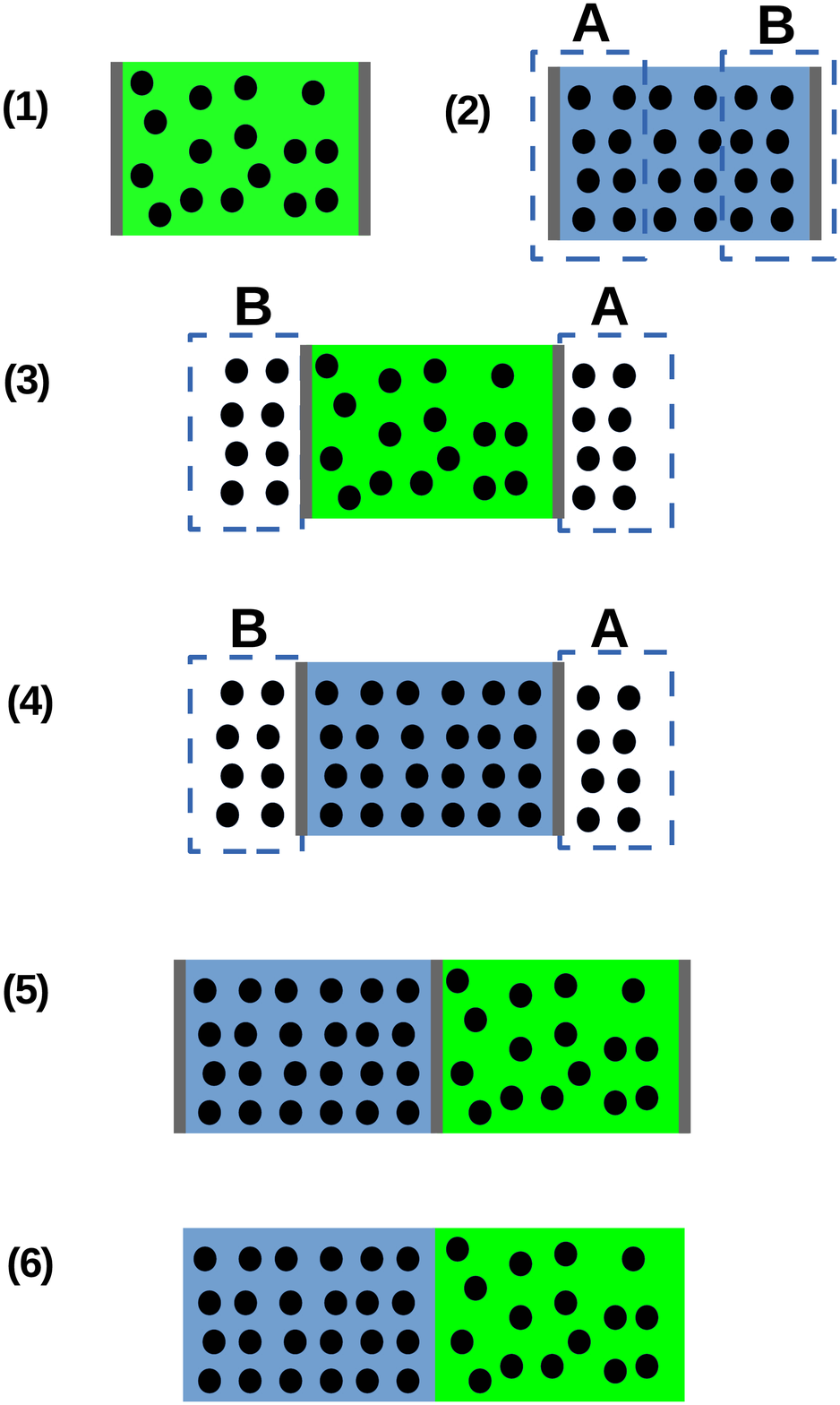}
\caption{\label{fig:TIscheme}
Sketch of the TI scheme to obtain $\gamma_{\rm cl}$. Particles in the
blue and green simulation boxes represent the crystal and the liquid
phase, respectively. For details see text.}
\end{figure}
The interfacial free energy $\gamma_{\rm cl}$ is the excess free
energy required to form an interface between a crystal and a liquid
at coexistence. Since it is an excess free energy, given by the
difference between the free energies of a final and an initial state,
it can be directly computed via TI. While the initial state is given
by two independent systems, namely a bulk crystal and a bulk liquid at
coexistence temperature and pressure, the final state is an inhomogeneous
system where the liquid and the crystal phase are separated from each
other by two interfaces. When constructing the reversible thermodynamic
path between the initial and the final state, the bulk regions of the
two phases should be perturbed as little as possible and no stress should
be generated in the crystal when it gets in contact with the liquid.

A crucial step in our TI scheme is the introduction of an extremely
short-ranged flat wall which is modelled by a repulsive Gaussian potential
and placed at the boundaries of the simulation box in $z$ direction.
The range of the particle interactions with this wall is chosen such
that it is about $1000-10000$ times less than the typical size of
the particles. Thus, the Gaussian wall only prevents the liquid and
crystalline particles from crossing the boundaries of their simulation
boxes but does not change the thermodynamic properties of the bulk system
in any manner.  Due to the extremely short-range nature of the wall, a
very small time-step is needed to integrate the equations of motion in a
molecular dynamics simulation. However, one can use a multiple time-step
molecular dynamics algorithm to tackle this issue~\cite{frenkel-smit02}
and since only a few particles interact with the short-ranged wall, the
additional computational overhead in implementing this algorithm is small
(i.e.~the simulations with Gaussian walls are less than a factor of two
slower than corresponding simulations without these walls).

To ensure minimal perturbation of the crystal when the two phases are
brought together, it is essential that the liquid is already ordered into
crystalline layers near the interface, compatible with the actual crystal
structure. In our TI scheme, this is achieved by introducing a structured
solid wall, consisting of particles frozen into a configuration adopted
in an actual simulation of the crystal phase. Interactions between the
system and this structured wall were modelled by the same interaction
potential as that between the system particles. Such a structured solid
wall leads to formation of interfacial layers in the liquid which are
more compatible with the actual crystal structure than if the structured
wall consisted of particles fixed to an ideal crystal position as in
Refs.~\cite{davidchack-laird2000,davidchack-laird03}.

The starting points of our TI scheme are the simulations of a bulk crystal
and a bulk liquid under coexistence conditions (see Sec.~\ref{sec_sim}).
Both the liquid and the crystal are simulated in cubic boxes of identical
dimensions ($L_{\rm x}\times L_{\rm y}\times L_{\rm z}$) but at their
respective coexistence densities, assuming periodic boundary conditions
in all spatial directions.  The goal is to join simulation boxes with
the liquid and the crystal phase along the $z$ direction and thereby
create two independent crystal-liquid interfaces in equilibrium with the
bulk phases (with the total dimension $L_{\rm x}\times L_{\rm y}\times
2L_{\rm z}$).  Our TI scheme consists of the following sequence of steps
to carry out this transformation (see Fig.~\ref{fig:TIscheme}):

{\it{Step 1}}: Insert an extremely short-ranged Gaussian flat wall at
both ends of the liquid simulation box along the $z$ direction [sketch
(1) in Fig.~\ref{fig:TIscheme}], while keeping the periodic boundary
conditions intact.  The wall should be strong enough to prevent
the liquid particles from crossing the boundaries and be extremely
short-ranged such that the interfacial free energy of the liquid in
contact with this wall is negligible as compared to $\gamma_{\rm cl}$
(in our case, typically less than the statistical errors reported in
previous works~\cite{davidchack-laird03}).

{\it{Step 2}}: Identical flat walls are inserted at the boundaries of the
simulation box containing the crystal [(2) in Fig.~\ref{fig:TIscheme}]
such that the distance of the closest crystalline layer from the
flat wall is identical for both the left and right boundaries of the
simulation cell.

{\it Step 3}: Two solid walls are constructed from the left and right
parts of the crystal simulation cell near the two boundaries, containing
2-3 layers of crystalline particles.  Then, these walls are gradually
attached to the liquid simulation cell at the appropriate ends, with the
flat walls still switched on, to prevent particles from crossing the
boundaries. As shown in Fig.~\ref{fig:TIscheme}, the wall constructed
from the left part of the crystal simulation cell is attached to the
right end of the liquid simulation cell.  Similarly, the wall made from
the right side of the crystalline simulation cell is attached to the
left side of the liquid phase.  At the same time, the periodic boundary
conditions of the liquid simulation cell in $z$ direction are gradually
switched off. In this respect, this step is unlike the procedure adopted
in previous TI schemes, where the boundary conditions are kept intact
while switching on the ``cleaving'' walls.

{\it Step 4}: The same frozen-in parts of the crystal simulation cell
are gradually introduced to the appropriate boundaries of the simulation
box containing the actual crystalline phase, in presence of the flat
walls, and simultaneously the periodic boundary conditions in the $z$
direction are turned off [cf.~(4) in Fig.~\ref{fig:TIscheme}].

{\it Step 5}: The liquid and crystal systems are brought together and
simultaneously the frozen-in crystalline walls are removed, in presence of
the flat walls. This is accomplished by joining appropriate ends of the
crystal and liquid simulation cells, where the flat walls are located.
Doing this entails juxtaposing the solid wall in contact with one phase
to be on top of the other phase. However, both phases interacted only
with their respective solid walls. Since the periodic boundary conditions
for the individual phases were already switched off, one needs only to
switch on interactions between the two phases across the ends connected
directly and also across the other end via gradually turning on the
periodic boundary conditions [cf.~(5) in Fig.~\ref{fig:TIscheme}].  This
procedure is different from that followed by the ``cleaving potential'' or
``cleaving walls'' approach, where the liquid and crystal are first joined
together and only in the next step, the ``cleaving walls'' are removed.

At the end of step 5, the system consists of crystal and liquid phases
separated by two interfaces, whose position coincides with the position
of the flat walls. Since, the crystal and liquid particles cannot cross
the boundaries of their respective simulation cells, 
the position of the crystal-liquid interface
is tied to the position of the flat walls. If we proceed in a reverse
direction from the end of step 5, and then retrace each prior step,
we will end up exactly from where the transformation started, i.e.~the
liquid and crystal phases at coexistence in separate simulation cells.

{\it Step 6}: Finally, in the last step, the extremely short-ranged
Gaussian walls are gradually removed. When the the barrier imposed by
the walls becomes weaker, the particles can cross the boundaries of
their simulation cells. As a result the interfaces can move leading to
hysteresis in the thermodynamic integration path.  However, our scheme
is successful in tackling the hysteresis arising from the mobility of
the crystal-liquid interface, because the contribution of the last step
to $\gamma_{\rm cl}$ is negligible (less than the combined statistical
errors of steps 3, 4 and 5), on account of the extremely short-ranged
flat walls. This ensures that any residual hysteresis in the last step
is also reduced to a negligible amount, thus having no impact on the
accuracy of the estimates. This is a more desirable way to circumvent
the problem rather than the cumbersome approach of carrying out several
forward and reverse TI runs of various durations and choosing the path
with the least hysteresis~\cite{davidchack-laird03,davidchack2010}.

In our TI calculations, we carry out the sequence of transformations
outlined above by directly modifying the interaction potential of the flat
walls and the frozen-in crystalline walls. This approach
is similar to our earlier work on the determination of interfacial free
energies of liquid or crystal phases in contact with flat or structured
walls~\cite{benjamin-horbach2012,benjamin-horbach2013}. 

In all the steps of our TI scheme,
the parameter $\lambda$ is coupled to the interaction potentials in a 
non-linear rather than a linear manner~\cite{frenkel-smit02}. This is dictated
by the need to obtain smooth thermodynamic integrands leading to an accurate
numerical determination of the associated integrals. The particular choice
for the $\lambda$-parameterizations adopted in the following are however, not 
unique, and other variants have been tried by us with identical results.
In the following,
we describe the specific manner in which the sequence of steps outlined
above is carried out by coupling a parameter $\lambda$ to the interaction
potential between the system and the walls. 

{\it Steps 1 and 2}: Here, the interaction of a particle $i$ with an
extremely short-ranged flat wall is modelled by a Gaussian potential,
\begin{equation}
 \label{eq:flatgwall}
u_{\text{fw}}(z_{iw})=
a \exp\left[-\left(\frac{z_{iw}}{b}\right)^{2}\right] \, ,
\end{equation}
with $z_{iw}$ the distance of the particle from the wall in $z$-direction.
The parameters $a$ and $b$ control the height of the potential barrier
and the range of the potential, respectively. A suitable choice for $a$
is about 20-25\,$k_{\rm B}T$ (with $k_{\rm B}$ the Boltzmann constant
and $T$ the temperature of the system), which is sufficient to make the
walls impenetrable for particles near the boundaries. The parameter $b$
is set to around 0.0001\,$\sigma$--0.001\,$\sigma$, i.e.~a factor of
1000-10000 smaller than the typical size of the LJ particles.  This choice
of $b$ provides that the free energy contribution due to the Gaussian
wall is much less than the statistical error bars in the most precise
calculations of $\gamma_{\rm cl}$. The Gaussian flat walls are placed
at the boundaries of the simulation cell at $z=0$ and $z=L_{\rm z}$.

A parameter $\lambda$ is coupled to the flat wall as follows,
\begin{equation}
\begin{split}
\label{eq:uwallsubst}
u_{\text{fw}}(\lambda,z_{iw})=& \lambda^{2} u_{\text{fw}}(z_{iw})
\end{split}
\end{equation}
Thus, at $\lambda=0$, the Gaussian wall is zero, and as $\lambda$
increases the wall becomes more and more impenetrable and finally fully
impenetrable at $\lambda=1$.

The $\lambda-$dependent Hamiltonian for steps 1 and 2 takes the form
\begin{equation}
\begin{split}
H_{1,2}(\lambda) = \sum_{i=1}^{N}
\frac{{{\bf p}_{i}^{2}}}{2m}
+ {U_{\text{pp}}^{\text{c(l)}}} 
+ \lambda^{2}{U_{\text{fw}}^{\text{c(l)}}}\, ,
\end{split}
\label{eq:hamilt_step12}
\end{equation}
where, $H_{\rm 1}$ represents interaction of the flat wall with the liquid
particles and $H_{\rm 2}$ those of the flat wall with the crystalline particles.
In Eq.~(\ref{eq:hamilt_step12}), ${\bf p}_{i}$ and $m$ represent the momentum and mass 
of particle $i$ with
all particles have the same mass. The contribution from
the particle-particle interactions to the potential energy is given by
${U_{\text{pp}}^{\text{c(l)}}}=\sum_{i=1}^{N^{\text{c(l)}}}\sum_{j=i+1}^{N^{\text{c(l)}}}u(r_{ij})$
with $r_{ij}$ the distance between two
particles $i$ and $j$.  The potential energy due
to the interactions of particles with the flat wall is
${U_{\text{fw}}^{\text{c(l)}}}=\sum_{i=1}^{N^{\text{c(l)}}}u_{\text{fw}}(z_{\text{iw}})$,
where the superscript $c(l)$ refers to particles in the crystal (liquid)
phase and $N^\text{c}$ and $N^\text{l}$ are the total number of liquid
and crystal particles, respectively. Dimensions of the simulation cells
containing individual crystal and liquid phases are kept identical and
as $\rho_{\rm l}<\rho_{\rm c}$, $N_{\rm c}<N_{\rm l}$.

\begin{figure}
\includegraphics[width=3.2in]{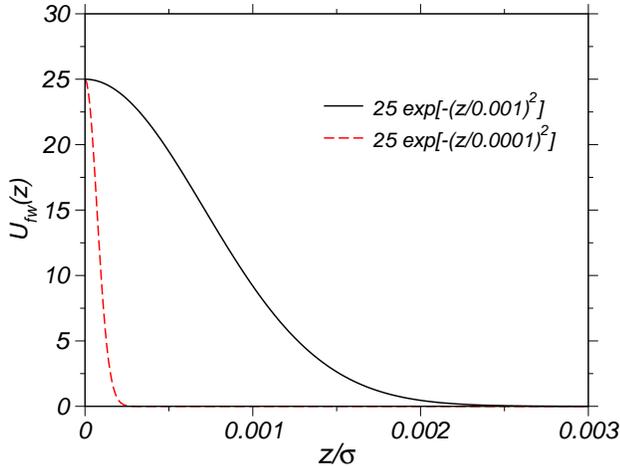}
\caption{\label{fig:gflatwall}
Gaussian flat wall potential for $b=10^{-3}\,\sigma$ (black solid line)
and $b=10^{-4}$ (red dashed line); in both cases, the barrier height is
set to $a=25\,\epsilon$, as used for $T=1.0$. For $T=0.617$ and $T=1.5$,
the latter parameter was set to $a=15\,\epsilon$ and $a=35\,\epsilon$,
respectively.}
\end{figure}
The free energy differences, $\Delta F_{1}$ and $\Delta F_{2}$, in steps
1 and 2 are obtained via the following integration over the parameter
$\lambda$,
\begin{equation}
\Delta F_{1,2} =  
\int_{0}^{1} \left\langle 
\frac{\partial H_{1,2}}{\partial \lambda} 
\right\rangle  d\lambda \nonumber 
= \int_{0}^{1} \left\langle 2 \lambda {U_{\text{fw}}^{\text{c(l)}}} 
\right\rangle d\lambda \,.
\end{equation}

Since the crystal can translate along the $z$ direction, it is not
guaranteed that the crystal is symmetrically positioned with respect to
the left and right boundaries in $z$ direction. This creates a problem
in the subsequent steps.  Since the solid walls inserted in steps 3
and 4 are made of frozen-in crystalline layers near the two ends of the
crystal simulation cell, such asymmetry in the position of the crystal
will lead to two dissimilar solid walls being made to come in contact
with the liquid and crystal phases on either side. This would lead to
an ordering in the liquid phase which is not compatible with the actual
crystal structure. Similarly, two walls with their outermost layers at
different distances from the two sides of the crystal phase might lead
to stresses in the crystal.

To avoid this, the two innermost layers of the crystalline phase were
frozen such that the crystalline phase was symmetric with respect to
the two boundaries. This ensured that the solid walls generated from
such a symmetric crystal were similar and their outermost layers were
at the same average distance from the two boundaries. This is similar to
what was done in the ``cleaving wall'' approach~\cite{davidchack-laird03}.  
Freezing the innermost
layers did not have any additional side effects on the crystal-liquid
interface, as was verified by carrying out simulations with both larger
and shorter $L_{\rm z}$.

{\it Steps 3 and 4}: The $\lambda$-dependent Hamiltonian for steps 3
and 4 is chosen as follows,
\begin{equation}
\begin{split}
H_{3,4}(\lambda) =  
\sum_{\text{i=1}}^{N} \frac{{\bf p}_i^{2}}{2m}
+ {{U}_{\text{pp}}^{\text{c(l)}}}
+ (1-\lambda)^{3}{{U^{\ast}}_{\text{pp}}^{\text{c(l)}}}\\
+ \lambda^{10}{U_{\text{pw}}^{\text{c(l)}}}
+ {U_{\text{fw}}^{\text{c(l)}}}
\end{split}
\label{eq:hamilt_step34}
\end{equation}
where $H_3$ includes the interactions of the structured wall with
the liquid particles and $H_4$ those of the structured wall with the
crystalline particles.  ${{U^{\ast}}_{\text{pp}}^{\text{c(l)}}}$
represents interactions between particles through
the periodic boundaries along the $z$ direction, while
${U_{\text{pp}}^{\text{c(l)}}}$ corresponds to direct interactions.
${U_{\text{pw}}^{\text{c(l)}}}=\sum_{i=1}^{N^{\text{c(l)}}}
\sum_{j=1}^{N^{\text{w}}}u_{\text{pw}}(r_{{ij}})$,
where $N^{\text{w}}$ is the total number of particles comprising the
frozen walls and interactions between the system and the structured
wall particles is denoted by $u_{\text{pw}}$.  For $u_{\text{pw}}$
the potential given by Eqs.~(\ref{eq:ur1}) and (\ref{eq:ur2}) was
used, replacing the parameter $\epsilon$ by $\epsilon_{\rm pw}$. In
step 3, $\epsilon_{\rm pw}$ varied between $0.54\,\epsilon$ and
$1.0\,\epsilon$ (depending on the considered temperatures), while in
step 4, $\epsilon_{\rm pw}=1.0\,\epsilon$ was set.  These choices of
$\epsilon_{\rm pw}$ ensured that the bulk densities of the liquid and
crystal phases remained unperturbed throughout the integration path.

The free energy differences for steps 3 and 4 are obtained as
\begin{eqnarray}
\Delta F_{3,4} & = & 
\int_{0}^{1} \left\langle \frac{\partial H_{3,4}}{\partial \lambda} 
\right\rangle  d\lambda \nonumber \\
& = & - \int_{0}^{1} \left\langle 3(1-\lambda)^{2} 
{{U^{\ast}}_{\text{pp}}^{\text{c(l)}}} \right \rangle d\lambda 
\nonumber \\ 
& & + \int_{0}^{1} \left\langle 10 \lambda^{9}
{U_{\text{pw}}^{\text{c(l)}}} \right \rangle d\lambda \; .
\end{eqnarray}

Since we directly modify the interaction potential between the structured
solid wall and the system particles and the system particles cannot cross
the boundaries of their simulation boxes due to the repulsive Gaussian
flat walls, there is no need of adopting any ``corrugated cleaving plane''
as was introduced in Refs.~\cite{davidchack-laird03,davidchack2010}.

{\textit Step 5}: In this step, the Hamiltonian is given by
\begin{equation}
\begin{split}
H_{5}(\lambda) =  
\sum_{\text{i=1}}^{N_{p}} 
\frac{{\bf p}_{\text{i}}^{2}}{2m_{i}}
+ {{{U}_{\text{pp}}^{\text{c(l)}}}}
+ \lambda^{5}{U_{\text{pp}}^{\text{c+l}}}\\
+ (1-\lambda)^{5}{U_{\text{pw}}^{\text{c(l)}}}
+ {U_{\text{fw}}^{\text{c(l)}}}
\end{split}
\label{eq:hamilt_step5}
\end{equation}
where ${U_{\text{pp}}^{c+l}}= \sum_{i=1}^{N^{\text{l}}}
\sum_{j=1}^{N^{\text{c}}}u(r_{{ij}})$ corresponds to interaction between
a liquid particle (with index $i$) and a crystalline particle (with
index $j$).

The free-energy difference for step 5 is
\begin{eqnarray}
\Delta F_{5} & = & 
\int_{0}^{1} \left[\left\langle \frac{\partial H_{5}}{\partial \lambda} 
\right\rangle \right] d\lambda \nonumber \\
& = & 5 \int_{0}^{1} \langle \lambda^{4} 
{U_{\text{pp}}^{\text{c+l}}} \rangle \, d\lambda \nonumber \\
& & - 5 \int_{0}^{1} \langle (1-\lambda)^{4} 
{U_{\text{pw}}^{\text{c(l)}}} \rangle \, d\lambda \,.
\end{eqnarray}
The interaction strength between the structured wall and the liquid
and crystal particles was kept at the same value of $\epsilon_{\rm pw}$
as in steps three and four, respectively.

{\textit Step 6}: In the last step, the Gaussian flat walls are gradually
switched off. In this case, the following $\lambda$ parametrization is
used for $u_{\rm fw}$,
\begin{equation}
\begin{split}
\label{eq:uwallfwstep6}
u_{\text{fw}}(\lambda,z_{iw})=&
(1-\lambda)^{2} u_{\text{fw}}(z_{\text{iw}}) \, ,
\end{split}
\end{equation}
and the $\lambda$-dependent Hamiltonian for this step takes the form,
\begin{equation}
\begin{split}
H_{6}(\lambda) = 
\sum_{i=1}^{N_{\rm p}} \frac{{{\bf p}_{i}^{2}}}{2m} 
+ {{{U}_{\text{pp}}^{\text{c(l)}}}}+ {U_{\text{pp}}^{\text{c+l}}}+
 {(1-\lambda)^2}{U_{\text{fw}}^{\text{c(l)}}}
\end{split}
\label{eq:hamilt_step6}
\end{equation}
and thus the free energy difference for step 6 is
\begin{equation}
\Delta F_{6} = 
\int_{0}^{1} \left\langle \frac{\partial H_{6}}{\partial \lambda} 
\right\rangle d\lambda
= \int_{0}^{1} \left \langle 2 (\lambda-1) 
{U_{\text{fw}}^{\text{c(l)}}} \right \rangle d\lambda
\label{eq:free_eng6}
\end{equation}

The interfacial free energy is finally obtained by summing over
the equilibrium free energy differences for the six sequential
transformations and the division by the total interfacial area $A$,
\begin{equation}
\gamma_{\rm cl} = \frac{{\Delta F}_{1}+{\Delta F}_{2} 
                      + {\Delta F}_{3}+{\Delta F}_{4} 
                      + {\Delta F}_{5}+{\Delta F}_{6}}{A} \, ,
\end{equation}
with $A=2L_{\rm x}L_{\rm y}$ (note that the factor 2 takes into account
that due to the periodic boundary conditions there are two independent
planar crystal-liquid interfaces, cf.~Fig.~\ref{fig:TIscheme}).

The TI scheme proposed in this work contains six steps as opposed to
the four steps involved in the ``cleaving potential'' or ``cleaving
walls'' scheme. However, contributions from steps one, two and six
are negligible on account of the very short-ranged flat walls. As our
results in Sec.~IV indicate, these contributions are less than the total
statistical errors from steps three, four and five. Therefore, one could
start with independent liquid and crystal phases in contact with Gaussian
flat walls and end with two crystal-liquid interfaces in equilibrium
with the bulk phase and in presence of such short-ranged walls and be
still able to obtain reliable estimates for $\gamma_{\rm cl}$.  We also
want to point out that such short-ranged Gaussian walls do not suppress
capillary wave fluctuations~\cite{morris2002,rozas2011,schmitz2014}
but only prevent the movement of the two crystal-liquid interfaces.

\subsection{Simulations}
\label{sec_sim}
Molecular dynamics (MD) computer simulations are performed at constant
particle number $N$, constant volume $V$, and constant temperature $T$.
To keep the temperature constant, the velocities of the particles were
drawn from a Maxwell-Boltzmann distribution at the desired temperature
every $200$ time steps. To integrate the equations of motion, the
velocity form of the Verlet algorithm~\cite{allen-tildesley87} is used.
We use a multiple-time step scheme~\cite{frenkel-smit02} to take into
account the short-range forces due to the Gaussian wall. For this purpose,
a smaller time step of $\Delta t_{\rm small}=0.00025\,\tau$ was used in
conjunction with a larger time-step of $\Delta t_{\rm large}=0.004\,\tau$,
where $\tau=\sqrt{(m\sigma^{2}/\epsilon)}$.

\begin{figure}
\includegraphics[width=3.0in]{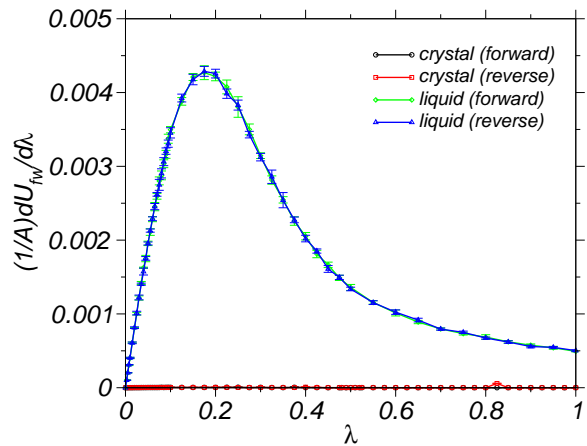}
\caption{\label{fig:step1ti}
Thermodynamic integrand for steps 1 and 2 for the liquid and crystal
phases, respectively, at $T=1.0$ and the (100) orientation of the
crystal. Error bars in this figure and subsequent ones represent one
standard deviation.}
\end{figure}
The density $\rho=N/V$ of the crystal and liquid at coexistence,
corresponding to the modified LJ potential considered in this work, were
taken from Ref.~\cite{davidchack-laird03}.  We consider three different
coexistence temperatures at $T=0.617$, 1.00, and 1.50.  The coexistence
densities of the crystal and liquid, $\rho_{\rm l}$ and $\rho_{\rm c}$,
respectively, at these temperatures are as follows: $\rho_{l}=0.8282$
and $\rho_{c}=0.945$ at $T=0.617$ and a coexistence pressure $P=-0.02$,
$\rho_{l}=0.923$ and $\rho_{c}=1.0044$ at $T=1.00$ and a coexistence
pressure $P=4.95$ and $\rho_{\rm l}=1.003$ and $\rho_{\rm c}=1.074$ at
$T=1.50$ and a coexistence pressure $P=12.95$. All the pressure values are
in reduced units, $\sigma^{3}/k_{B}T$.  We have verified the coexistence
densities by carrying out simulations of crystal-liquid interfaces and
checking the densities of the crystal and liquid in the bulk. Independent
simulations of bulk crystal and liquid were also carried out at these
densities and it was confirmed that the coexistence pressures in both the
bulk liquid and bulk crystal were identical within the statistical error.

The parameters $a$ and $b$ appearing in the Gaussian flat wall
potential have to be chosen carefully such that the wall is extremely
short-ranged. At the same time, when it is fully applied no particles
should  cross the barrier imposed by this short-ranged wall. Taking
these two factors into account, $a$ was taken to be $15.0\epsilon$, $25.0\epsilon$ and
$35.0\epsilon$ at the co-existence temperatures $T=0.617$, $1.0$ and $1.5$,
respectively. At each temperature, $b$ was chosen to take the value
$0.001 \sigma$, which made the wall sufficiently short-ranged. Nevertheless,
we also tried another wall with an even shorter range i.e.~$b=0.0001\sigma$
but obtained identical results within the statistical error.

System sizes for the various coexistence temperatures and orientations of
the crystal were the same as in Ref.~\cite{davidchack-laird03}. Only at
$T=1.0$ and for the (100) orientation of the crystal-liquid interface,
simulations were carried out for various system sizes, corresponding
to several combinations of lateral ($L_{\rm x}$, $L_{\rm y}$) and
longitudinal ($2 L_{\rm z}$) system sizes. The dimensions of the
simulation boxes are specified in Table I. For the considered system
sizes, the total number of particles varied from $N=6000$ to about
$N=32000$.

To generate initial configurations, the liquid and crystal
phases were equilibrated at the coexistence temperatures and their
respective coexistence densities. Dimensions of the liquid and crystal
simulation cells were identical and since $\rho_{\rm l}<\rho_{\rm c}$,
correspondingly less number of particles were contained in the liquid
simulation cell. After each phase was equilibrated for around $200000$
time steps (here and later on, in multiples of the larger time-step $\Delta t_{\rm large}$), 
the TI simulations were started.

For the TI calculations, independent runs were carried out at various
values of $\lambda$ between $0$ and $1$.  For the various TI steps, the
total number of intervals between $\lambda=0$ and $\lambda=1$ varied
from $45$ to $50$, which was sufficient to produce smooth thermodynamic
integrands.  At each value of $\lambda$, i.e.~$\lambda_{i}$, the system
was equilibrated at $\lambda=0$ for $10000$ time steps and then $\lambda$
was continuously increased until $\lambda_{i}$ was reached. The number
of time-steps to carry out this switch varied from $90000$ to $150000$
time-steps for the various system sizes and various orientations. After
the final value $\lambda_{i}$ was reached, the system was further
equilibrated for $500000-750000$ time-steps.  Then the production
runs were carried out over a period ranging from $250000$ to $500000$
time-steps. Statistical errors were determined by dividing the production
runs into $5$ blocks and then obtaining the standard deviation between
these $5$ samples.

To calculate the free energy difference from the obtained thermodynamic
integrands for steps 3, 4 and 5, we did a cubic spline interpolation of
the bare data with 100 intervals between $\lambda=0$ and $\lambda=1$
and then used the Simpson rule to numerically calculate the integral.
For steps 1, 2 and 6, the numerical integration was carried out over
the bare data using the trapezoidal rule.

We have carried out simulations in both the forward and reverse directions
to detect any residual hysteresis in the TI path. The initial state for the
reverse TI simulations were taken from the final state of the forward TI path.
The final values for
$\gamma_{\rm cl}$ reported in the next section correspond to a mean
of the free energy differences obtained from the forward and reverse
TI simulations.

\begin{figure}
\includegraphics[width=3.0in]{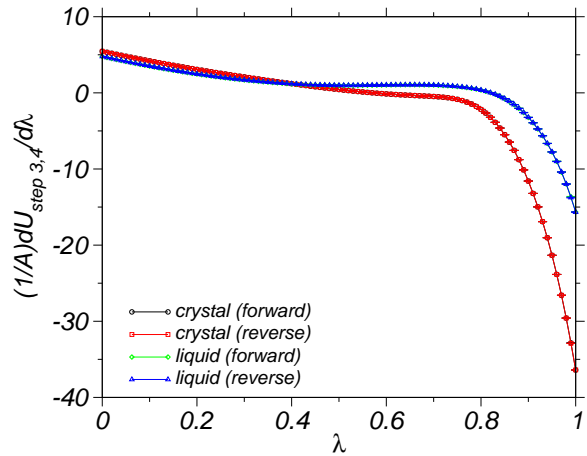}
\caption{\label{fig:step2ti}
Thermodynamic integrand corresponding to steps 3 and 4, for the liquid
and (100) orientation of the crystal, respectively, at the temperature
$T=1.0$.}
\end{figure}
\section{Results}
\label{sec_res}
\begin{figure}
\includegraphics[width=3.0in]{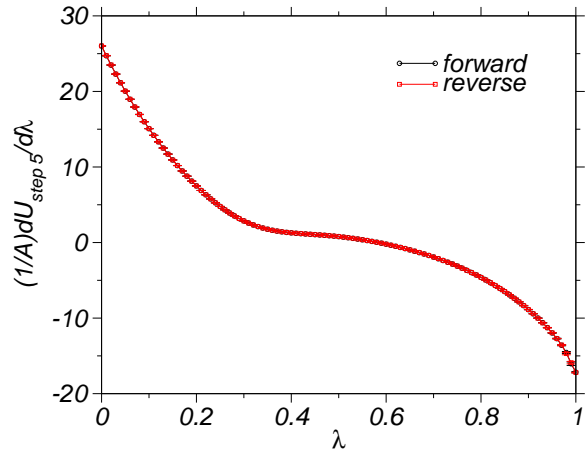}
\caption{\label{fig:step3ti}
Thermodynamic integrand for step 5 at $T=1.0$, bringing the (100)
orientation of the crystal in contact with the liquid.}
\end{figure}

\begin{figure}
\includegraphics[width=3.0in]{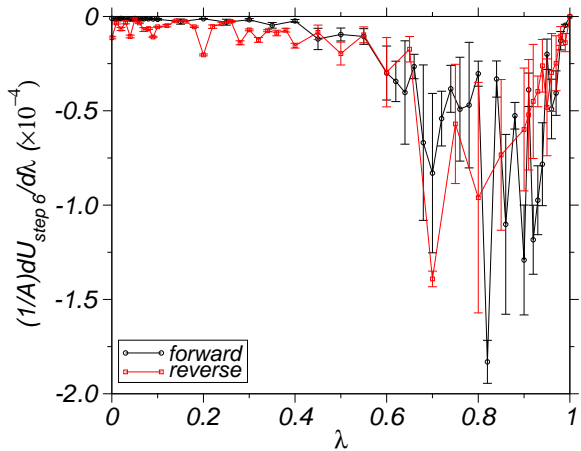}
\caption{\label{fig:step4ti}
Thermodynamic integrand for the last step, corresponding to switching
off the flat wall.}
\end{figure}

In Figs.~\ref{fig:step1ti}, \ref{fig:step2ti}, \ref{fig:step3ti}, and
\ref{fig:step4ti}, the thermodynamic integrands corresponding to the
six steps are plotted for the (100) orientation of the crystal-liquid
interface at $T=1.0$. Here, the system size of the final state with two
interfaces is $14.263 \sigma \times 14.263 \sigma \times 31.695 \sigma$.
In all figures we show the thermodynamic integrands for both the forward
(increasing $\lambda$ from $0$ to $1$) and reverse (decreasing $\lambda$
from $1$ to $0$) processes. The final values of the interfacial free
energy for the various cases are reported in Table 1 and the errors 
include both the statistical error and the residual hysteresis between 
the forward and reverse paths.

In Fig.~\ref{fig:step1ti}, we show the thermodynamic integrands for
inserting the Gaussian flat wall (and removing for the reverse process)
into the crystal and liquid phases, respectively.
Good overlap of the forward and reverse thermodynamic integrands
indicates, as expected, the lack of hysteresis in our TI scheme.
As can be clearly seen in the figure, the area under the thermodynamic
integrand corresponding to the crystal in contact with the flat wall is
negligible as compared to the same situation for the liquid. On account
of the extremely short range of the flat walls, very few particles
actually interact with them leading to a negligible contribution to
the free energy difference.  Since the crystal is placed symmetrically
with respect to the two ends of the simulation cell, the flat walls
are located in the middle of two crystalline layers.  Therefore, their
position coincides with a minimum of the crystal density leading to a
negligible value for the free energy difference as compared to the liquid.

The interfacial free energy of the liquid in contact with the flat wall
is $0.0018k_{B}T$ at $T=1.0$ when $b=0.001\sigma$, which is of the order of
$<0.35\%$ of the final value of $\gamma_{\rm cl}$.  In case of the crystal,
the contribution of this flat wall varies from $\sim 10^{-7}\,k_{\rm
B}T$ for the (111) orientation to $10^{-6} k_{\rm B}T$ for the (100)
orientation and $10^{-5} k_{\rm B}T$ for the (110) orientation. For this choice of $b$,
the combined contribution of step 1 and step 2 was lower than the statistical
errors in the most precise estimates of $\gamma_{\rm cl}$.

In the next two steps, liquid and crystalline phases were separately
brought into contact with structured solid walls in presence of the
Gaussian flat wall and simultaneously the periodic boundary conditions
were switched off.  Thermodynamic integrands corresponding to steps 3
and 4 are shown in Fig.~\ref{fig:step2ti}. In step 3, the interaction
strength $\epsilon_{\rm pw}=1.0\,\epsilon$ was chosen at $T=0.617$,
while at the other temperatures $\epsilon_{\rm pw}=0.54\,\epsilon$.
For the interaction of the crystal with the structured wall, at
all temperatures $\epsilon_{\rm pw}$ was set to $1.0\,\epsilon$.
As the $\lambda$ parameter is gradually increased, the liquid begins
to form ordered layers near the interface due to interactions with
the solid wall. This process was the source of some hysteresis
in the ``cleaving walls''~\cite{davidchack-laird03} and ``cleaving
potentials''~\cite{broughton-gilmer86} scheme, though it could be avoided
by equilibrating the samples for longer times.

The smooth variation of the thermodynamic integrands corresponding to
steps 3 and 4 (see Fig.~\ref{fig:step2ti}) along with small error bars
throughout the integration path indicates that our system was well
equilibrated and not trapped in some metastable state. The excellent
overlap between the thermodynamic integrands for the forward and reverse
processes also confirms the reversibility of our TI scheme.

In the next step, the liquid and crystal were joined together, while
the interactions of the liquid and crystal phases with the frozen walls
were switched off.  As Fig.~\ref{fig:step3ti} shows, the integrands
are smooth leading to an accurate numerical determination of the
integral. The forward and reverse thermodynamic integrands are also
in very good agreement. In the ``cleaving potential'' and ``cleaving
walls'' approach, the liquid and crystal phases were first brought
together and then the walls were removed. However, in our scheme these
steps are combined into a single step. Nevertheless, we carried out TI
simulations where our single step scheme was divided into two steps:
first bringing the two phases together and in the next step removing
the structured walls, while varying the parameter $\lambda$ in exactly
the same manner as specified in Eq.~\ref{eq:hamilt_step5}.  However,
both approaches led to identical results.

\begin{figure}
\includegraphics[width=3.0in]{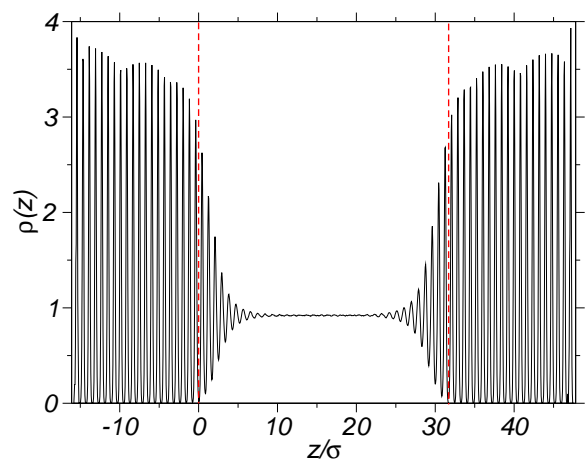}
\caption{\label{fig:dens_step4} 
Density profile of the (100) orientation of the crystal in contact with
the liquid and separated by a short-ranged flat wall. The temperature
is $T=1.0$. Red vertical lines correspond to the positions of the flat
walls. The system size is $14.263\,\sigma \times 14.263\,\sigma \times
31.695\sigma$.}
\end{figure}
\begin{figure}
\includegraphics[width=3.0in]{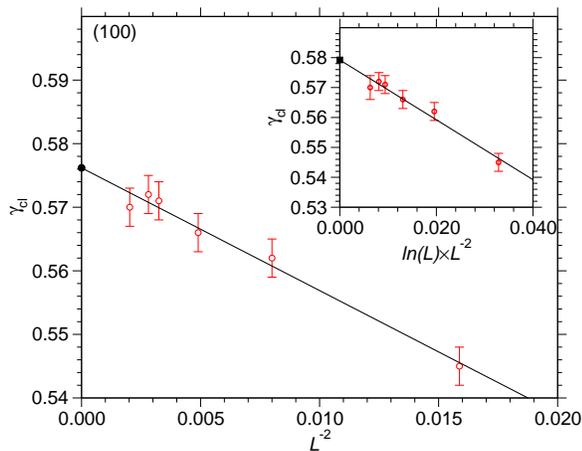}
\caption{\label{fig:gam_vs_l2inv}
$\gamma_{\rm cl}$ vs. $1/L^{2}$, where $L=L_{\rm x}=L_{\rm y}$,
corresponding to the (100) orientation of the crystal-liquid interface
with the longitudinal dimension $L_{\rm z}=31.695\,\sigma$. For the same
parameters, the inset shows $\gamma_{\rm cl}$ vs.~$\ln(L)/L^{2}$. Solid
symbols correspond to the $\gamma_{\rm cl}$ values in the thermodynamic
limit.}
\end{figure}
The final step involves removing the flat walls. However, this process
is no longer reversible after the barrier posed by the flat wall becomes
weaker and the crystal-liquid interfaces can move.  This is reflected in
the hysteresis between the forward and reverse thermodynamic integrands,
as shown in Fig.~\ref{fig:step4ti}. However, the contribution of
this step is negligible owing to the short-range nature of the flat
walls. Figure~\ref{fig:step4ti} clearly shows that the area under
both the forward and reverse thermodynamic integrands is much smaller,
in magnitude, as compared to step 1.  As a result, one does not need to
follow the ``cleaving walls'' approach of carrying out several forward and
reverse simulation runs to estimate the path with the least hysteresis.
In fact, for the (100) orientations of the crystal-liquid interface
both forward and reverse simulations predict a magnitude less than
$10^{-4}\,k_{\rm B}T$ for this step. This contribution will be negative
since the flat wall is purely repulsive and we are removing it (see
Eq.~\ref{eq:free_eng6} and Fig.~\ref{fig:step4ti}).  Similarly negligible
contributions are obtained for the other two orientations as well.

To understand why ${{\Delta F}_6}/A$ would be negligible, in
Fig.~\ref{fig:dens_step4} we plot the density profile corresponding to the
(100) orientation of the crystal-liquid interfaces in equilibrium with
their bulk phases. If the minima of the density profile coincides with the
flat wall position, one will obtain a negligible contribution as in step
2. On the other hand if a maxima coincides, the result would be larger,
though due to a finite barrier imposed by the repulsive Gaussian walls,
the probability of a maxima in the density profile coinciding with the
wall position is small.

In the first step, the free energy difference per unit area was around
$0.0018k_{B}T$ (at $T=1.0$). Since, we observe that the maxima in the
density profile is about 3 times that of the bulk liquid, it is clear
that the maximum possible magnitude for ${{\Delta F}_6}/A$ would be 3
times the contribution of ${{\Delta F}_1}/A$ and therefore would still
be significantly smaller than $\gamma_{\rm cl}$. Since the interfaces
can move by simultaneous melting and freezing, the average density of
the inhomogeneous system near the positions of the flat wall would be
much smaller. Therefore, the actual free energy difference in the last
step would be much less than this maximum possible value.

\begin{table*}[htbp]
\centering
\begin{tabular}{|c|c|c|c|c|c|}
\hline
\cline{1-5}
 $\text{Temperature}$&$\text{Orientation}$&$\text{System Size}$&$\gamma_{\rm cl}$(TI)&Cleaving Wall&Non-Eq. Work\\
\hline \hline
$1.000$&$100$&$7.924 \times 7.924 \times 31.695$&$0.545\pm0.003$&---&---\\
$1.000$&$100$&$11.093 \times 11.093 \times 31.695$&$0.562\pm0.004$&---&---\\
$1.000$&$100$&$14.263 \times 14.263 \times 31.695$&$0.566\pm0.003$&$0.562\pm0.006$&---\\
$1.000$&$100$&$19.017 \times 19.017 \times 20.602$&$0.572\pm0.003$&---&---\\
$1.000$&$100$&$17.432 \times 17.432 \times 31.695$&$0.571\pm0.003$&---&---\\
$1.000$&$100$&$19.017 \times 19.017 \times 31.695$&$0.572\pm0.003$&---&---\\
$1.000$&$100$&$19.017 \times 19.017 \times 41.204$&$0.569\pm0.004$&---&---\\
$1.000$&$100$&$22.187 \times 22.187 \times 31.695$&$0.570\pm0.003$&---&---\\
$1.000$&$100$&$ \infty $&$0.576^{\rm a},0.579^{\rm b}$&---&---\\
$1.000$&$110$&$14.263 \times 13.447 \times 26.894$&$0.545\pm0.003$&$0.543\pm0.006$&---\\
$1.000$&$111$&$11.646 \times 12.327 \times 32.939$&$0.515\pm0.006$&$0.508\pm0.008$&---\\
$0.617$&$100$&$14.559 \times 14.559 \times 32.352$&$0.372\pm0.005$&$0.371\pm0.003$&$0.371\pm0.004$\\
$0.617$&$110$&$14.559 \times 13.726 \times 27.452$&$0.357\pm0.003$&$0.360\pm0.003$&$0.361\pm0.003$\\
$0.617$&$111$&$11.887 \times 12.582 \times 33.622$&$0.344\pm0.006$&$0.347\pm0.003$&$0.354\pm0.003$\\
$1.500$&$100$&$13.951 \times 13.951 \times 31.001$&$0.866\pm0.005$&$0.84\pm0.02$&---\\
$1.500$&$110$&$13.951 \times 13.153 \times 26.306$&$0.785\pm0.006$&$0.82\pm0.02$&---\\
$1.500$&$111$&$11.391 \times 12.057 \times 32.218$&$0.774\pm0.007$&$0.75\pm0.03$&---\\
\hline
\end{tabular}
\caption{Interfacial free energy $\gamma_{\rm cl}$ for
different system sizes corresponding to various orientations
of the crystal-liquid interface, three different co-existence
temperatures and different system sizes. For comparison, data from
the cleaving wall~\cite{davidchack-laird03} and non-equilibrium work
apporaches~\cite{musong2006} are also shown.  Dimensions ($L_{\rm x }
\times L_{\rm y} \times L_{\rm z}$) are in units of $\sigma^{3}$. At
$T=1.0$ and for the (100) orientation of the crystal-liquid interface,
the interfacial free energy in the thermodynamic limit is extrapolated
from the  values of $\gamma_{\rm cl}$ at the various system sizes using
the (a) $L^{-2}$ and (b) $\ln(L)L^{-2}$ scalings (see text).}
\end{table*}

In principle, one can make the Gaussian flat walls as short-ranged as
possible to reduce their contribution even further. However, reducing
the range involves a concomitant increase in computational cost due to
the need for a very short time-step in the multiple-time step scheme. A
slightly more efficient way to reduce the hysteresis, would be to use
a Gaussian flat wall with a range long enough that one can use the
regular single time-step velocity-verlet algorithm and at the same
time sufficiently short-ranged, so as not to affect the bulk density
of the phases. One can carry out steps 1-5 of the TI scheme with this
flat wall, with no computational overhead involved in a multiple-time
step scheme. In the next step, one can first transform this flat wall
to an extremely short-ranged wall. The parametrization for this step,
which we call step 6a, will be
\begin{equation}
\begin{split}
\label{eq:uwallfwstep6a}
u_{\text{fw}}(\lambda,z)=&
 a\exp(-[z/b(\lambda)]^{2}) \, ,
\end{split}
\end{equation}
where $b(\lambda) = (1-\lambda)b^{\prime}$ with $b^{\prime}=0.001\sigma$
and $\lambda$ varying from $0$ to $0.9$.  This step will be subject
to minimal hysteresis as the flat walls are still present to prevent
the particles from crossing boundaries of their simulation cells. The
average position of the crystal-liquid interface would oscillate around
the position of the flat walls.  In the next step (step 6b) one can
remove the flat walls with the same parametrization as in Eq.~(8).

We have carried out simulations by breaking step six into the two steps
as specified here and obtained results in perfect agreement with the
earlier method, since the flat wall chosen for our simulations was
already extremely short-ranged. However, this latter approach of using
an extremely short-ranged wall only in the final step, will be useful
for long-ranged potentials, where the computational overhead of using a
multiple time-step scheme for all the six steps might become significant.

The flat Gaussian walls can, in principle, be adapted to any continuous
potential. Only the prefactor $a$, which sets the height of the
barrier needs to be modified to adapt the barrier height at different
temperatures. The value for the parameter $b$ would be suitable for most
potentials, and  in cases where $\gamma_{\rm cl}$ has a small magnitude, the 
approach of breaking down step six into two steps would be also useful.

In Table I, we report the final values of $\gamma_{\rm cl}$ for different
orientations, system sizes and temperatures. We also report data from the
cleaving wall method~\cite{davidchack-laird03} and a non-equilibrium work
approach~\cite{musong2006}. Clearly, our results are in good agreement
with both the methods.
 
While $\gamma_{\rm cl}$ can be computed by various molecular simulation
techniques, often there is disagreement between them which is greater
than the statistical errors. For example, values for the hard-sphere
interfacial free energy predicted by the ``cleaving wall'' approach
is more than $10\%$ less than the one computed using the capillary
fluctuation method or the tethered Monte Carlo approach.  In a recent
work, Schmitz {\it et al.}~\cite{schmitz2014} suggested systematic
errors arising out of finite size effects to be a possible source of
disagreement between the various methods.  They identified the mechanism
of finite size corrections and proposed a theoretical tool to obtain
reliable estimates for interfacial tensions in the thermodynamic limit.

In a three-dimensional system, if a planar interface is described by a 
lateral dimension $L=L_{\rm x}=L_{\rm y}$ and a longitudinal dimension $L_{\rm z}$,
the leading finite size corrections to the interfacial free energy in the 
thermodynamic limit, $\gamma_{\rm \infty}$, are described by the following 
formula \cite{schmitz2014,binder82}:
\begin{equation}
\gamma_{\rm L,L_{\rm z}} =
\gamma_{\rm \infty} -a\frac{\ln L_{\rm z}}{L^2} + b\frac{\ln L}{L^{2}} 
+ \frac{c}{L^{2}}\,
\label{eq:finitesize}
\end{equation}
where $a$, $b$ and $c$ are constants. To study the finite size
corrections in our system, we carried out simulations for various lateral
dimensions of the crystal-liquid interface for the (100) orientation of
the crystal-liquid interface at $T=1.0$.

In Eq.~(\ref{eq:finitesize}), the second term is associated with
the translational entropy of the interface due to movement of the
crystal-liquid interface. Since, the flat walls constrain the movement
of the crystal-liquid interface this term is negligible in our case
and can be neglected.  In Fig.~\ref{fig:gam_vs_l2inv}, we plot the
interfacial free energy for the (100) interface as a function of
$1/L^{\rm 2}$ and $\ln(L)/L^{\rm 2}$ (in the inset), at the longitudinal 
system size, $L_{\rm z}=31.695\sigma$. A linear extrapolation of the data 
provides $\gamma_{\rm cl}$ in the thermodynamic limit and for the $1/L^{\rm 2}$
scaling we obtain a value of $0.576$, while for the $\ln(L)/L^{\rm 2}$
scaling a value of $0.579$ is obtained.  Compared to the estimated value
of $\gamma_{\rm cl}$ at the lateral dimension studied by Davidchack and
Laird~\cite{davidchack-laird03}, viz.~$14.263\sigma \times 14.263\sigma$,
these values are larger by about $2\%$.

\section{Conclusion}
\label{sec_conc}
We have obtained the crystal-liquid interfacial free energy for the
Lennard-Jones potential via a novel thermodynamic integration scheme. A
crucial lacunae of previous thermodynamic integration schemes pertaining
to the movement of the crystal liquid interface has been overcome by the
help of extremely short-ranged Gaussian walls.  Another feature of our
scheme is the use of frozen-in crystalline layers to induce ordering
in the liquid and thereby merge the crystal and liquid in a smooth
manner. Our results are in good agreement with previous methods based on
the cleaving method~\cite{davidchack-laird03} as well as a non-equilibrium
work approach~\cite{musong2006}.  Using the finite-size scaling for the
(100) orientation, we have extrapolated the finite-size data to obtain
$\gamma_{\rm cl}$ in the thermodynamic limit.

Aside from our TI scheme with flat walls, other schemes such as the
ones proposed by Schilling and Schmid~\cite{schilling-schmid2009} or by
Grochola~\cite{grochola2004} could, in principle, be used to overcome the
problem related to movement of the crystal-liquid interface. However,
such schemes involve modifying the interaction potential of the entire
system, yielding large free energy differences and as a result the
statistical errors are of the same order of magnitude as $\gamma_{\rm
cl}$ itself. Therefore, such schemes have to be substantially modified
to compute $\gamma_{\rm cl}$.

In a recent work, $\gamma_{\rm cl}$ for the modified LJ potential has
been obtained using density functional theory~\cite{wangdft2013}, with
values in reasonable agreement with our data. The precise estimates for
the interfacial free energy could be used to further refine and validate
density functional theory approaches.

The thermodynamic integration scheme developed in this work can be used to
obtain the interfacial free energies of systems described by more complex
potentials or also of hard sphere systems, with only minor modification
of the coupling of the parameter $\lambda$ to the interaction potentials.
Work in this direction is the subject of forthcoming studies.

\begin{acknowledgments}
The authors acknowledge financial support by the German DFG SPP 1296,
Grant No.~HO 2231/6-3. J.~H.~acknowledges useful discussions with 
Kurt Binder, Fabian Schmitz and Peter Virnau.
\end{acknowledgments}

\end{document}